\documentclass{article} 
\usepackage[utf8]{inputenc}
\usepackage{url}
\usepackage{graphicx}
\usepackage{enumitem}
\usepackage{xcolor}
\usepackage{tabularx} 
\usepackage{booktabs} 
\usepackage{ragged2e} 
\usepackage{float}
\usepackage{listings}

\lstdefinelanguage{json}{
    basicstyle=\ttfamily\small,
    numbers=none,
    showstringspaces=false,
    breaklines=true,
    frame=single,
    backgroundcolor=\color{white},
    literate=
     *{0}{{{\color{blue}0}}}{1}
      {1}{{{\color{blue}1}}}{1}
      {2}{{{\color{blue}2}}}{1}
      {3}{{{\color{blue}3}}}{1}
      {4}{{{\color{blue}4}}}{1}
      {5}{{{\color{blue}5}}}{1}
      {6}{{{\color{blue}6}}}{1}
      {7}{{{\color{blue}7}}}{1}
      {8}{{{\color{blue}8}}}{1}
      {9}{{{\color{blue}9}}}{1}
      {:}{{{\color{black}:}}}{1}
      {,}{{{\color{black},}}}{1}
      {"}{{{\color{red}"}}}{1}
      {[}{{{\color{black}[}}}{1}
      {]}{{{\color{black}]}}}{1}
      {\{}{{{\color{black}\{}}}{1}
      {\}}{{{\color{black}\}}}}{1},
}

\begin{document}
\begin{titlepage}
    \centering
    \vspace*{2cm}  

    {\Huge\bfseries {\textbf{Knowledge Graph-Enhanced Multi-Agent
Infrastructure for coupling physical and digital
robotic environments(KG-MAS)}}\par}
    \vspace{1.5cm}

    {\Large \textbf{ A Research-Oriented Report}\par}
    \vspace{2cm}

    {\Large \textbf{Walid Abdela}\par}
    \vspace{1cm}

    {\large \textbf{Supervisors:} \textbf{Dr. Nesrine Hafiene, Prof. Flavien Balbo}\par}
    \vfill

    {\large JULY 2025\par}
\end{titlepage}

\thispagestyle{empty}  
\newpage  

\tableofcontents
\thispagestyle{empty}
\newpage

\begin{abstract}
The seamless integration of physical and digital environments in Cyber-Physical Systems(CPS), particularly within Industry 4.0, presents significant challenges stemming from system heterogeneity and complexity. Traditional approaches often rely on rigid, data-centric solutions like co-simulation frameworks or brittle point-to-point middleware bridges, which lack the semantic richness and flexibility required for intelligent, autonomous coordination. This report introduces the Knowledge Graph-Enhanced Multi-Agent Infrastructure(KG-MAS), as resolution in addressing such limitations. KG-MAS leverages a centralized Knowledge Graph (KG) as a dynamic, shared world model, providing a common semantic foundation for a Multi-Agent System(MAS). Autonomous agents, representing both physical and digital components, query this KG for decision-making and update it with real-time state information. The infrastructure features a model-driven architecture which facilitates the automatic generation of agents from semantic descriptions, thereby simplifying system extension and maintenance. By abstracting away underlying communication protocols and providing a unified, intelligent coordination mechanism, KG-MAS offers a robust, scalable, and flexible solution for coupling heterogeneous physical and digital robotic environments.
\end{abstract}
\thispagestyle{empty}
\newpage
\pagenumbering{arabic}
\setcounter{page}{1}
\section{Introduction} 
In sectors such as Industry 4.0, the seamless integration of digital and physical environments in the design of Cyber-Physical Systems (CPSs) has become increasingly challenging\cite{hafiene2025knowledge}. The complexity of these systems stems from their inherently heterogeneous nature, involving diverse communication protocols, varying constraints, and the need for synchronized coordination across physical and digital domains.

Traditional solutions in managing CPSs have typically treated physical and digital environments as separate entities, each operating within its own technological ecosystem. For instance, physical robotic platforms often utilize specific middleware frameworks such as Robot Operating System (ROS) with distinct communication protocols, while digital simulation environments operate under different paradigms with their own data models and interaction mechanisms. This separation creates a significant barrier for system designers who seek to develop and test robotic solutions.

The challenge is further compounded by the dynamic and unpredictable nature of physical environments, which contrasts sharply with digital or simulated environments that typically operate under fewer spatial and temporal constraints and are less affected by environmental noise. Additionally, maintaining a coherent global overview of the system becomes increasingly difficult when integrating both physical and simulated components. For instance, it is technically possible for a physical robot and a simulated robot to occupy the same virtual position within the system. However, this is physically impossible in the real world. Current methodologies for coupling these environments often rely on rigid domain-specific solutions that lack the flexibility to accommodate the evolving requirements of modern CPSs. Although existing cosimulation frameworks provide structured approaches to system integration through standards such as the High-Level Architecture (HLA) \cite{dahmann1997department} and the Functional Mock-up Interface (FMI)\cite{blochwitz2011functional}, they often do not address the need for dynamic knowledge management and real-time coordination between heterogeneous robotic platforms\cite{ndiaye2018simulation}.

Knowledge graphs can be viewed as a promising solution for representing complex, interconnected data in a structured, query-able manner\cite{hogan2021knowledge}. When combined with multi-agent systems, which provide a means for distributed coordination and decision-making, knowledge graphs can provide a unified representation of system states, environmental data, and inter-component communications. This combination offers the potential to bridge the gap between physical and digital environments by providing a common semantic foundation for information exchange and coordination.

The integration of multi-agent systems with knowledge graphs presents an opportunity to address the fundamental challenges of cyber-physical system coupling \cite{crooks2011integration}. Multi-agent systems inherently support distributed coordination mechanisms that can manage the complex interactions between system components, while knowledge graphs provide a standardized, domain-independent approach to knowledge representation that can accommodate the diverse data types and relationships present in modern robotic systems. Multi-agent programming environments, such as Hypermedea, provide the necessary infrastructure to implement such integrated solutions\cite{charpenay2022hypermedea}.

Contemporary cosimulation approaches have explored various coupling mechanisms, including sequential and parallel coupling strategies, with different time synchronization methods ranging from time-stepped to event-driven approaches \cite{hafner2013investigation, mindra2024cyber}. However, these frameworks often lack the semantic richness and dynamic adaptability required for complex robotic coordination scenarios. The need for more improvised approaches that can handle both the technical heterogeneity of robotic platforms and the semantic complexity of multi-domain coordination remains a significant challenge.

The remainder of this report is organized as follows. Section 2 presents a comprehensive review of the state of the art, explaining how different methodologies are used in coordinating heterogeneous CPS environments. In Section 3, the overall implementation of the proposed infrastructure is discussed, including how it was planned, designed, built and tested. Section 4 outlines avenues for future work, specifically the formalization of a FIPA-ACL-based coordination protocol and the incorporation of collision avoidance representations within the knowledge graph. Following this, Section 5 conducts a comparative analysis, evaluating the proposed KG-MAS solution against existing state-of-the-art methodologies using a set of key integration criteria. Finally, Section 6 concludes the paper by summarizing the contributions of KG-MAS for robust and flexible CPS integration.

\section{Related Work}
 Establishing a seamless integration between physical and digital components in CPSs have become a pronounced challenge in Industry 4.0. Such integration is critical for designing, testing, and deploying robotic applications efficiently and safely. This section discusses about the several distinct methodologies which have been developed to address this issue.

\subsection{Co-Simulation Frameworks}
At the most fundamental level, the problem of coupling disparate systems is addressed by co-simulation standards. These frameworks provide a structured methodology for orchestrating the joint execution of multiple, independent simulation units. The two most influential standards in this domain are the High-Level Architecture(HLA) and the Functional Mock-up Interface(FMI). HLA, defines a federation of simulation components that interact through a central RunTime Infrastructure (RTI). The RTI is responsible for managing data exchange and advancing simulation time in a coordinated manner, ensuring that all components share a consistent temporal view of the system\cite{dahmann1997department}. 

On the other hand, FMI provides a tool-independent standard for packaging simulation models into Functional Mock-up Units(FMUs). Each FMU is a black-box model with a standardized API, allowing a master algorithm to control its execution and link its inputs and outputs with other FMUs\cite{blochwitz2011functional}. These frameworks are powerful for achieving data-level interoperability. For example enabling a physics simulator and a network simulator to exchange information at synchronized time steps.

While powerful for synchronized data exchange, these frameworks primarily focus on the mechanics of coupling simulators. They neither inherently describe a persistent model of a physical asset nor do they typically incorporate the complexities of a CPS.

\subsection{Middleware Integration}
Software bridges focus creating direct translators between different communication protocols and middleware systems. The most prominent example in the robotics community is the \textbf{ros1\_bridge}, a software package designed to transparently forward messages between ROS1 and ROS2, automatically translating message formats where necessary\cite{macenski2020ros1bridge}. Similar bridges have been developed to connect ROS with other prevalent protocols, such as MQTT for IoT applications or DDS for real-time systems.

This approach is highly effective for solving a specific, point-to-point integration problem, allowing a legacy ROS1-based robot to communicate with a modern ROS2-based control system. However, this solution can be brittle; it requires a separate bridge for each pair of protocols, and the coordination logic must still be manually coded by the developer, who needs to be aware of the different systems involved. Moreover, this approach lacks a centralized world model and a high-level coordination framework, making it less scalable and flexible for complex systems with many heterogeneous components. 

\subsection{Digital twins}
The Digital Twin(DT) paradigm represents an evolution from session-based co-simulation to a persistent, synchronized virtual representation of a physical asset, process, or system\cite{grieves2014digital}. A DT architecture is characterized by a continuous, bidirectional flow of information between the physical object and its virtual counterpart. Sensors on the physical system feed real-time data to the digital model, ensuring it accurately reflects the state of the real world. In turn, the digital model can be used for monitoring, analysis, and especially for simulating ``what-if" scenarios by interacting with other virtual components. 

The results of these simulations can then be used to optimize operations or send commands back to the physical asset\cite{cimino2019digital}. In the context of robotics, a physical robot's digital twin can be placed in a fully simulated factory environment to test a new collaborative task before deploying the new configuration in the real world. The core contribution of the DT approach is its focus on maintaining a live, synchronized link that bridges the entire lifecycle of the system. However, the power of a DT is often limited by the expressiveness of its underlying data model, which may consist of raw database entries or proprietary CAD(Computer-Aided Design) formats.

\subsection{Ontologies}
Knowledge-based approaches employ formal ontologies and KGs to create a shared, machine-readable understanding of a system. An ontology provides a formal vocabulary to describe entities, their properties, and the relationships between them\cite{gruber1993translation}. By building a system model on a shared ontology, such as the Smart Applications REFerence (SAREF) ontology\cite{garcia2023etsi}, different agents and components can unambiguously interpret system data. Frameworks like RoboEarth and KnowRob have pioneered this approach, creating knowledge bases where robots can query for task information, learn from the experiences of other robots, and reason about their environment\cite{waibel2011roboearth, beetz2018knowrob}. 

In this paradigm a query is not just for a value, but for the position of a robot. A semantically rich request that enables more intelligent and flexible coordination. This semantic layer provides the memory for an integrated system, allowing components to understand the context and purpose of their actions.

\subsection{RAMI 4.0}
The Reference Architectural Model for Industrie 4.0(RAMI 4.0 — DIN SPEC 91345) serves as a robust framework for structuring Industry 4.0(I4.0) systems, enabling
seamless integration of cyber-physical production systems(CPPSs) and advancing digital transformation in manufacturing.\cite{platform2016rami}. It provides three distinct categories for managing the complexity of production environments which are Life Cycle and Value Stream, Hierarchy Levels, and Layers.

The Life Cycle and Value Stream dimensional(IEC 62890), covers the entire lifecycle of products and systems from design and production to maintenance and recycling, ensuring a holistic view of value creation. The Hierarchy Levels dimension(IEC 2264), organizes systems vertically from field devices like sensors to enterprise-level systems such as Enterprise Resource Planning(ERP), enabling seamless data flow across organizational scales. Finally, the Layers dimension comprised of six layers, \textbf{Asset; Integration; Communication; Information; Functional; and Business}, decomposes system functionalities into modular components fostering interoperability and structured data management \cite{bachhofner2021modularization, din2016rami}. Altogether, these categories provide a cohesive framework for navigating I4.0’s complexity.
\begin{figure}[h]
    \centering
    \includegraphics[width=1\linewidth]{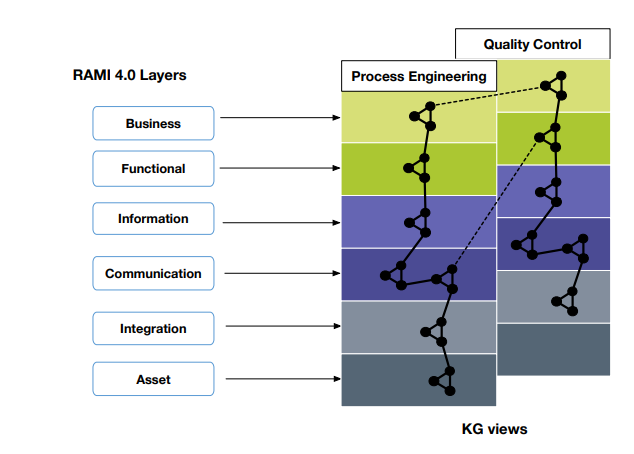}
    \caption{Modularized Knowledge Graph in Smart Manufacturing}
    \label{fig:modularized-kg}
\end{figure}
The proposed infrastructure will be considering the layered approach as it  fits to the nature of the project and also provides an intuitive way of organizing CPPSs. It allows the construction of modular knowledge graphs with incremental development, which is used in building use-case-specific views that aggregate into a comprehensive CPPS-wide KG unified by shared core concepts(Figure~\ref{fig:modularized-kg}).The six layers are defined as follows:
    
\begin{itemize}
  \item \textbf{Asset Layer:} Encompasses physical components, such as machines and tools, and their digital representations, forming the basis for data generation in CPPSs.
  
  \item \textbf{Integration Layer:} Connects physical assets to digital systems, facilitating data acquisition through sensors and actuators.
  
  \item \textbf{Communication Layer:} Enables standardized data exchange between components through protocols, ensuring interoperability across systems.
  
  \item \textbf{Information Layer:} Processes and aggregates raw data into meaningful insights, providing context for operational decisions.
  
  \item \textbf{Functional Layer:} Hosts application logic, such as AI-driven analytics or quality control algorithms, to optimize processes.
  
  \item \textbf{Business Layer:} Aligns production activities with strategic objectives and linking operational data to goals.\cite{bachhofner2021modularization, din2016rami}.
\end{itemize}

The layered approach extends beyond KG modularization to various I4.0 applications. Bader et al.\cite{bader2020knowledge} use the layers to classify I4.0 standards within a KG, mapping standards to specific layers to enhance interoperability. The Asset Administration Shell(AAS) leverages the layers to formalize asset semantics, improving system integration and asset management\cite{bader2019semantic}. In smart manufacturing, Pedone and Mezgár\cite{pedone2021interoperability} demonstrate how the layers enable data flow from IoT devices at the Asset and Integration Layers to AI-driven analytics at the Functional and Business Layers. Lin et al.\cite{lin2022digital} apply the layers to structure digital twins for predictive maintenance, integrating data across production lifecycles.

The preceding review of existing methodologies, from co-simulation standards to DTs, reveals their inherent limitations in providing the semantic flexibility and high-level coordination required for integrating CPSs. Consequently, this project proposes a different approach which revolves around a KG-driven, multi-agent infrastructure in solving the problem. Inspirations from the organizational principles of the RAMI 4.0 layered model were drawn in order to bring structure and modularize the system's knowledge base. The proceeding section will now detail the practical implementation of this infrastructure, demonstrating how this KG-centric model addresses the shortcomings of conventional approaches.

\section{Implementation}
The following sections discuss about the implementation details of the system by providing a general overview first and then focus on the Implementation strategy, its design, the development tools, and testing. The system is designed to handle the coupling of physical and digital robotic environments, with the help of autonomous agents facilitated by the Hypermedea framework and knowledge graphs.

\subsection{Overview}
\begin{figure}[h]
    \centering
    \includegraphics[width=1\linewidth]{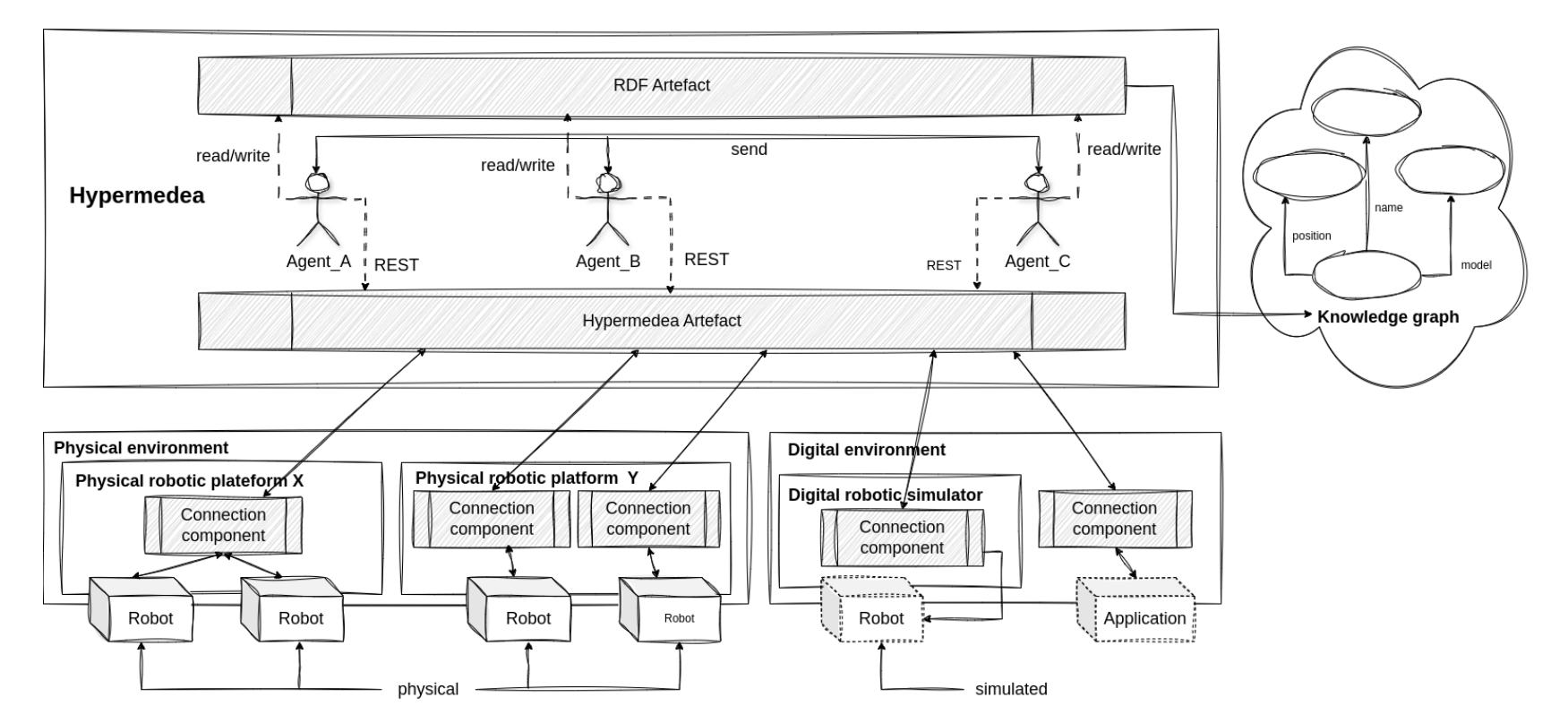}
    \caption{The Architecture of the proposed infrastructure}
    \label{fig:system-overview}
\end{figure}

The proposed infrastructure introduces an architectural paradigm that leverages a MAS enhanced by a centralized KG. The core objective is to create a unified ecosystem where physical and digital robotic components can be coupled and coordinated dynamically, regardless of their underlying technology. This is achieved by abstracting away the communication complexities and establishing a shared, real-time global knowledge-based model that all system components can reference for decision-making. The infrastructure thereby simplifies system design, enables robust and flexible coordination, and allows for the validation of simulated components in coherence with physical hardware before significant capital investment. The following are the key components that make up the infrastructure(Figure~\ref{fig:system-overview}):

\begin{itemize}
  \item \textbf{Hypermedea:} Multi-agent programming environment used for developing autonomous agents which represent the different components of a CPS. Based on JaCaMo (Java + Jason + Moise), it ensures the seamless coupling between the physical and digital environments via the coordination among the agents which is facilitated by artifacts. Artifacts are interfaces that an agent can use for interacting with different components. For example, an agent can use the Hypermedea Artifact which is composed of REST operations in order to communicate with the physical environment and can also use the RDF Artifact which is composed of SPARQL queries in order to communicate with the KG.
  
  \item \textbf{Knoweledge graph:} Common frame of reference used by agents which functions as a centralized, dynamic, and shared memory for the entire system. The KG stores a structured representation of the current state of the entire world, including both physical and digital entities.
  
  \item \textbf{Connection Component:} A translator that receives the command from the Hypermedea artifact and converts it into the final, native command that the specific device understands. The device can also communicate the information that it perceives using the Connection Component.
  
  \item \textbf{Physical/Digital Environments:} Represents the physical and digital robotic platforms where actions are executed via the commands received from the agents.
  
\end{itemize}

\subsection{Strategy}
The system's implementation was carried out in the following manner:

\subsubsection{Knowledge Graph Development(April-May)}

\begin{itemize}
     \item Defined a conceptual resolution for integrating knowledge graphs with Hypermedea agents.
     
    \item Designed and implemented a setup knowledge graph which is based on the representing the system's initial configuration.

    \item  Built a supplementary knowledge graph to store and manage information related to the state of the system.
\end{itemize}

\subsubsection{Generation of Autonomous Agents and Coordination Protocol Initialization(June)}
\begin{itemize}
    \item Implemented an agent creator component which is used to produce Hypermedea agents with the information available on the  knowledge graph.

    \item  Enhanced the agent creation process by generalizing it to  produce agents which support different models such as REST, HTTP, and COAP etc.

     \item  Designed the basic structure and layout of the coordiation protocol which is used in defining communications among agents. This ensures a synchronized flow of interactions and reliable coordination between agents controlling physical and simulated robots.

\end{itemize}

\subsection{System Design}
\subsubsection{Knowledge graphs}
The system is architected around the use of  knowledge graphs, which serve as a centralized, dynamic model for representing environmental data, robot states, and interactions between the physical and digital robotic environments. There are two types of knowledge graphs deployed within the system which are the following:
\begin{itemize}
  \item \textbf{System Setup:}  Contains all the necessary initial configuration in order to create autonomous agents.
  
  \item \textbf{System Data:} Stores information related to each specific robot, for example their states and positions.
  
\end{itemize}

\subsubsection{Revised RAMI 4.0}
A modified version of the RAMI 4.0 layered approach(Figure~\ref{fig:revised-rami}) was applied in order to cater to the implementation of the proposed infrastructure. In the original version, a system can be organized, as seen previously, into six distinct layers which are the \textbf{Asset}, \textbf{Integration}, \textbf{Communication}, \textbf{Information}, \textbf{Functional}, and \textbf{Business} layers. However, this variant will only be considering the following layers:
\begin{figure}[h]
    \centering
    \includegraphics[width=0.75\linewidth]{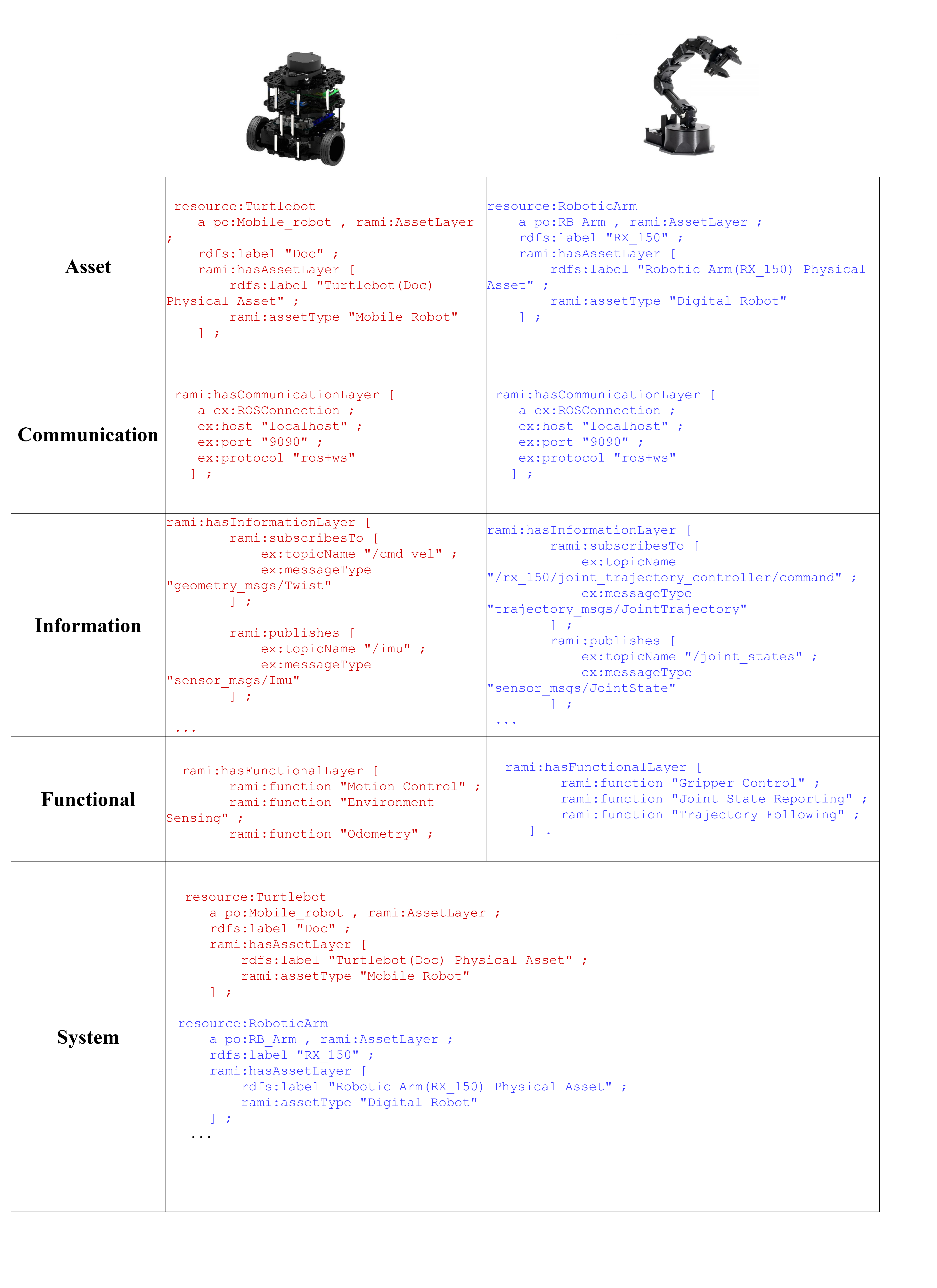}
    \caption{Revised layered approach of the RAMI 4.0}
    \label{fig:revised-rami}
\end{figure}
\begin{itemize}
\item \textbf{Asset:} Defines the entities themselves, such as a physical or digital robot.
    
\item \textbf{Communication:} Specifies the communication protocols associated with each asset.
   
\item \textbf{Information:} Represents the specific data streams the asset consumes(subscribes to) and produces(publishes).

\item \textbf{Functional:} Describes the high-level capabilities and functions of the asset in semantic terms. It moves beyond raw data to describe the asset's purpose and role within the system.

\item \textbf{System:} An aggregator that combines the information of each asset from the previous layers. This in turn facilitates the generation of simulations based on the complete information of the system.  
\end{itemize}

Figure~\ref{fig:revised-rami} demonstrates an example of this layered approach, showing how both the physical TurtleBot and the digital Robotic Arm are defined in the KG. For instance, the Asset layer defines the resources Turtlebot and RoboticArm as a \textbf{physical Mobile\_Robot} and a \textbf{digital Robotic\_Arm} asset respectively. The Communication layer then provides the necessary connection details, such as the\textbf{ ros+ws} protocol and the \textbf{localhost:9090} endpoint for both of the entities. Building on this, the Information layer specifies the exact\textbf{ ROS} topics each asset uses, like \textbf{/cmd\_vel} for the TurtleBot's movement commands and \textbf{/joint\_states} for the arm to report its status. Finally, the Functional layer describes their operability in
an abstract manner with high-level terms such as Motion Control for the TurtleBot and Gripper Control for the arm. This entire set of descriptions is then aggregated under a single system entity, forming a complete and self-contained definition for each component.

\subsubsection{Generation of Agents}
\begin{figure}[h]
    \centering
    \includegraphics[width=1\linewidth]{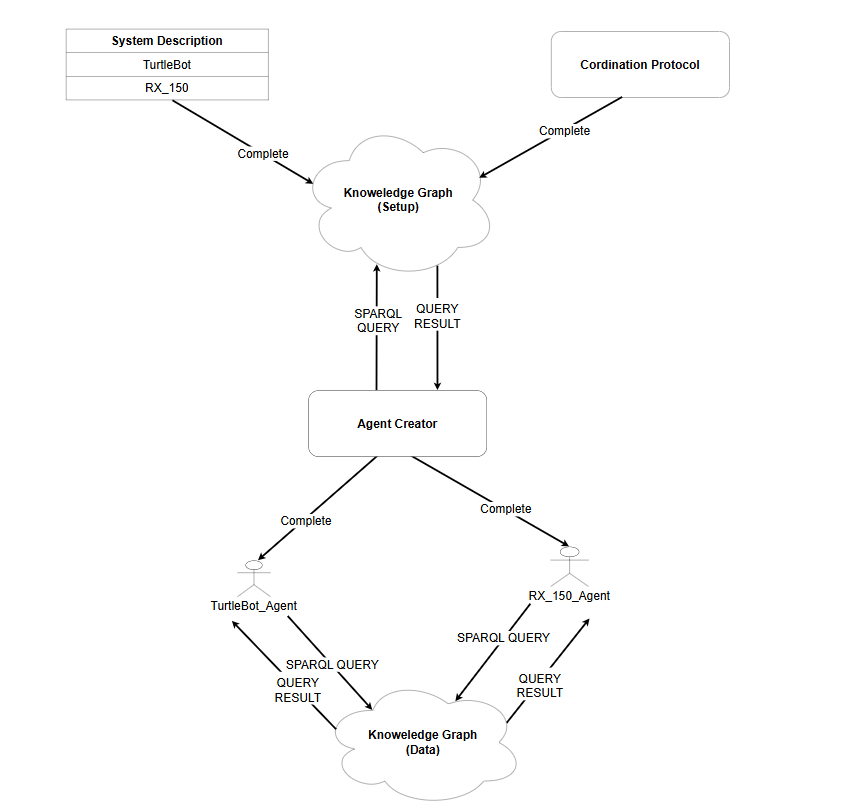}
    \caption{Agent generation process}
    \label{fig:agent-creation}
\end{figure}
The agent generation process is designed to be highly flexible, enabling the Agent Creator component to generate protocol-specific agents(REST, MQTT,and HTTP etc.) based on the preferences of the system designer. The workflow, as illustrated in Figure~\ref{fig:agent-creation}, transforms system configurations into fully operational, autonomous agents. This provides a minimalist effort approach for the system designer to automatically generate agents instead of hard-coding specific agent files. Consequently, this facilitates the work of the designer to focus on implementing other features of the system.

The process begins by populating the system's setup KG. This KG serves as the foundational repository, containing a complete description of all the system's entities and coordination protocol, where the information is organized using the revised RAMI 4.0 layered approach. Once the setup KG is populated, the Agent Creator queries this structured repository to obtain the necessary blueprints for building each agent. For example, it retrieves an entity's asset type, its designated communication protocol, and its functional capabilities. By dynamically building each agent's code based on these query results, the system eliminates the need for manual configuration and hard-coding, which in turn reduces development overhead and enhances system adaptability.

Following their successful instantiation, the newly created agents operate within a dynamic environment where they interact with a supplementary KG. The information stored in this KG is chosen to represent the real-time, operational state of the system, as opposed to the static configuration of the first. This dynamic data includes the current state and position of each robot and the status of ongoing tasks and interactions between agents. This clear distinction between a static configuration KG and a dynamic data KG allows the system to preserve its initial setup while effectively managing the constantly changing state of its operational environment.

\subsection{System development tools}
The following tools were used in the implementation of the system:

\begin{itemize}
\item \textbf{Java-17:} Used for implementing the agent creator component.
   
\item \textbf{Protégé:} Ontology modeling tool for defining domain-specific ontologies.
  
\item \textbf{ROS 1/2 \& Gazebo:}  Simulated robotic arm(RX150) and physical mobile robot(Turtlebot) environments.
   
\item \textbf{GraphDB:} RDF triplestore for managing the KG. Enables SPARQL queries and semantic reasoning over heterogeneous data.

\end{itemize}

\subsubsection{Testing}
\begin{figure}[h]
    \centering
    \includegraphics[width=0.75\linewidth]{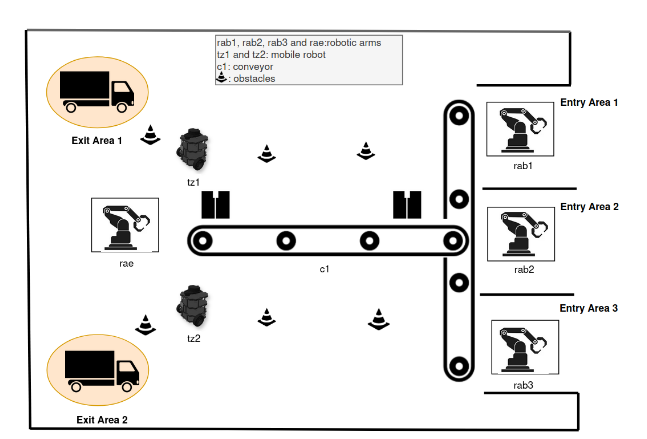}
    \caption{Motivating Scenario}
    \label{fig:motivating-scenario}
\end{figure}

The folllowing test setup, based on the warehouse scenario(Figure~\ref{fig:motivating-scenario}) was conducted in order to validate the proposed solution:

\subsubsection{Components}

\begin{itemize}
\item \textbf{Simulated Environment:} A Trossen Interbotix ReactorX150(RX\_150) robotic arm simulated in Gazebo.

\item \textbf{Physical Environment:} A Turtlebot3 Burger mobile robot deployed in the warehouse. 
    
\item \textbf{Multi-Agent System:} Agents for both the simulated and physical robots were initialized via the Agent Creation process and then subsequently implemented within the Hypermedea framework.
  
\item \textbf{Knowledge Graphs:} A knowledge graph containing the system's initial configuration was developed. Moreover, an additional knowledge graph was used for storing the information related to the different robots through respective agents.
   
\item \textbf{GraphDB:} RDF triplestore for managing the KG. Enables SPARQL queries and semantic reasoning over heterogeneous data.

\end{itemize}

\subsubsection{Results}

\begin{itemize}
\item \textbf{Minimum Code Adaptability:} The agent creator workflow successfully generated agents based on the available information on the system's knowledge graph without requiring them to be implemented from scratch by the system designer.

\item \textbf{Supplementary knowledge graph's dynamic information retrieval/Update:} Agents demonstrated the ability to dynamically retrieve/update on the supplementary knowledge graph with real-time information from the robots. 

\end{itemize}

\section{Future Work}
\subsection{Coordination Protocol}
The Coordination Protocol delineates the mechanisms by which heterogeneous agents exchange messages in order to synchronize their actions and interoperate seamlessly.  Leveraging the FIPA-ACL (Foundation for Intelligent Physical Agents - Agent Communication Language) standard, this protocol specifies message types, their formats, and the sequences of exchanges required for specific tasks. Furthermore, the protocol is structured to be stored within a knowledge graph, which will subsequently facilitate the automatic generation of the code that determines the communicative behavior of each agent within the system. 

As communication is facilitated through FIPA-ACL messages, each message comprises several essential fields. The performative field indicates the type of communicative act being performed, such as request, inform, confirm, refuse, or failure. The sender field identifies the agent dispatching the message, while the receiver field specifies the intended recipient agent. The content field encapsulates the actual information or request being communicated, typically structured in a JSON-like syntax to enhance clarity and ensure interoperability among heterogeneous agents. For example, a request message to initiate a task might be structured as:

\begin{lstlisting}[language=json, caption=Task initiation message, captionpos=b]
{"task": "move_pallet", "from": "P1", "to": "P2"}
\end{lstlisting}
\newpage
Additionally, an inform message reporting a state change could appear as:
\begin{lstlisting}[language=json, caption=State change message, captionpos=b]
{"event": "pallet_placed"}
\end{lstlisting}

\begin{figure}[h]
    \centering
    \includegraphics[width=1\linewidth]{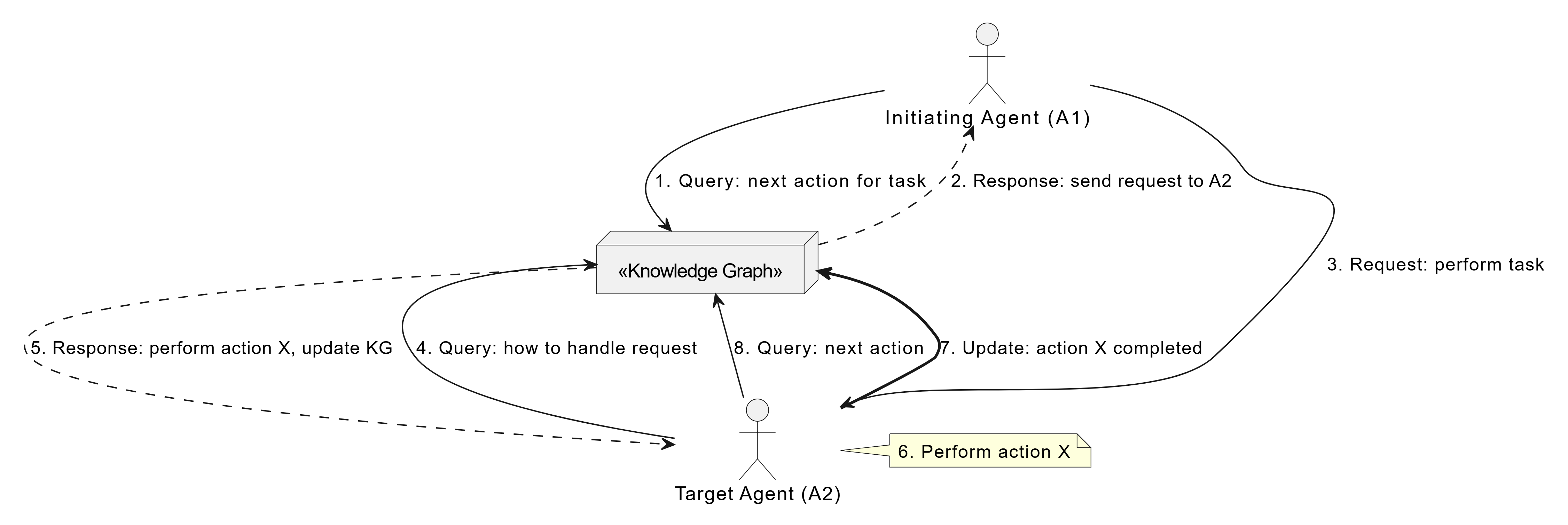}
    \caption{A high level Description of the Coordination protocol}
    \label{fig:coordination-protocol}
\end{figure}

A representative example of the interactions between agents using the cordination protocol(Figure~\ref{fig:coordination-protocol}), is as follows: 
\begin{itemize}
\item \textbf{A1 Queries the KG for the Next Action:} Agent A1 sends a FIPA-ACL message to the KG to request the next action for a task. The message has a Performative of request, Sender as A1, Receiver as KG, and Content like \texttt{\{"query": "next\_action", "task": "coordinate\_task"\}}.

\item \textbf{KG Responds to A1 with Instructions:} The KG processes A1’s query and responds with a FIPA-ACL message instructing A1 to send a request to Agent A2. The message uses a Performative of inform, Sender as KG, Receiver as A1, and Content such as \texttt{\{"action": "send\_request", "to": "A2", "task": "perform\_task"\}}. 

\item \textbf{A1 Sends a Request to A2:} A1 sends a FIPA-ACL message directly to A2, requesting it to perform a specific task. The message has a Performative of request, Sender as A1, Receiver as A2, and Content like \texttt{\{"task": "perform\_task", "details": \{"param1": "value1", "param2": "value2"\}\}} (e.g., \texttt{\{"task": "move\_pallet", "from": "P1", "to": "P2"\}}).

\item \textbf{KG Instructs A2 to Perform an Action:} Upon receiving A1’s request, A2 sends a FIPA-ACL message to the KG to seek guidance on how to handle it. The message uses a Performative of request, Sender as A2, Receiver as KG, and Content like \texttt{\{"query": "handle\_request", "from": "A1", "task": "perform\_task"\}}.

\item \textbf{A2 Performs the Action:} A2 executes the action specified by the KG, such as moving a pallet from point P1 to P2 or completing a subtask. This step involves no message exchange but represents A2 performing a physical or internal operation based on the instructions received.

\item \textbf{A2 Updates the KG:} After completing the action, A2 sends a FIPA-ACL message to the KG to report the result. The message has a Performative of inform, Sender as A2, Receiver as KG, and Content like \texttt{\{"event": "action\_completed", "action": "X", "status": "success"\}} (e.g., \texttt{\{"event": "pallet\_placed"\}}).

\item \textbf{A2 Queries the KG for the Next Step (optional):} If further actions are required, A2 sends another FIPA-ACL message to the KG to request the next step.
\end{itemize}

\subsection{Knowledge graph enhancements by implementing collision/obstacle detection}
The integration of obstacle detection and collision avoidance mechanisms into the knowledge graphs will be explored to enhance the reliability of the multi-agent infrastructure. The current system uses the knowledge graph to represent environmental data, and robot states, but it does not explicitly address obstacle detection or collision prevention. By implementing ontologies to include representations of both static obstacles(walls or fixed machinery) and dynamic obstacles(humans or robots), agents can access environmental information for safer decision-making. This could involve adding nodes for obstacles and relationships indicating potential collision risks, and incorporating real-time data from perception systems to anticipate future states based on trajectories. Such enhancements will ensure the safety and reliability of robotic systems in dynamic, unpredictable physical environments, making the system more suitable for real-world Industry 4.0 applications.

\section{Comparative Analysis}
 Coupling physical and digital robotic environments can be multifaceted, and as such, a variety of solutions have emerged, each tackling the problem from a different angle. The proposed KG-MAS solution by Hafiene et al.\cite{hafiene2025knowledge}, offers a  holistic approach by synthesizing concepts from several of these domains. To understand its specific contributions and trade-offs, this section provides a comparative analysis against the prominent state-of-the-art methodologies, evaluating them against a set of key criteria essential for robust and flexible system integration.

 \subsection{Comparison Criteria}
 The following criteria have been selected to evaluate each methodology's effectiveness in solving the integration problem:
 
\begin{itemize}
\item \textbf{Coordination \& Control:} Assesses the mechanism provided for orchestrating the behavior of different system components. It evaluates whether control is centralized or decentralized, and if it supports autonomous, goal-oriented decision-making.

\item \textbf{World Model Representation:} Examines how the state of the system and its environment is stored, shared, and updated. Key aspects include whether the model is merely data-centric or semantic, dynamic, and centralized.

\item \textbf{Heterogenity:} Evaluates the methodology's ability to manage and abstract the technical differences between various systems, particularly their underlying communication protocols.

\item \textbf{Scalability \& Felixibility:} Assesses the ease with which the system can be extended. It considers how new components can be added and how the system adapts to evolving requirements without major architectural overhauls.

\end{itemize}

\subsection{Analysis}
\begin{table}[H]
\centering
\caption{Comparative Analysis}
\label{tab:comparison_cosim_middleware_dt}
\small
\begin{tabularx}{\textwidth}{@{} l >{\RaggedRight}X >{\RaggedRight}X >{\RaggedRight}X @{}}
\toprule
\textbf{Criterion} & 
\textbf{Co-simulation Standards \newline (Dahmann et al.\cite{dahmann1997department}; Blochwitz et al.\cite{blochwitz2011functional})} & 
\textbf{Middleware Bridges \newline (Macenski et al.\cite{macenski2020ros1bridge})} & 
\textbf{Digital Twin (DT) \newline (Grieves \cite{grieves2014digital}; Cimino et al.\cite{cimino2019digital})} \\
\midrule

\textbf{Coordination \& Control} & 
Centralized master or federated control & 
None. Control logic is entirely user-defined and implemented elsewhere. & 
Not inherently defined \\
\addlinespace

\textbf{World Model Representation} & 
Data-centric model & 
None & 
High-fidelity digital replica \\
\addlinespace

\textbf{Handling Heterogeneity} & 
Manages heterogeneous simulators & 
Solves a specific protocol pair directly (e.g., ROS1 to ROS2) & 
Handled via custom data adapters and ingestion pipelines \\
\addlinespace

\textbf{Scalability \& Flexibility} &
Standardized but can be complex to reconfigure & 
A new bridge is needed for each new protocol pair. & 
Varies by implementation \\
\bottomrule
\end{tabularx}
\end{table}

The comparative analysis, as presented in Table~\ref{tab:comparison_cosim_middleware_dt}, evaluates the previously discussed methodologies for integration in CPSs, focusing on \textit{Co-simulation Standards}, \textit{Middleware Bridges}, and \textit{Digital Twins}. When viewed alongside the proposed KG-MAS solution, this analysis reveals how KG-MAS synthesizes the strengths of these paradigms while mitigating their inherent weaknesses.

For instance, while Co-simulation and DTs provide structured data exchange mechanisms and high-fidelity system replicas, their \textbf{World Model Representation} remains fundamentally data-centric. In contrast, KG-MAS employs a semantic KG that captures not only data but also the meaning and interrelationships among system entities. This enables a far richer and more dynamic \textbf{Coordination \& Control} mechanism, where autonomous agents can act goal-oriented. An ability not inherently present in DTs and only rigidly defined in centralized or federated control within co-simulation environments.

Furthermore, KG-MAS's approach to \textbf{handling heterogeneity} proves more scalable and robust compared to Middleware Bridges. Middleware solutions like the \textbf{ros1\_bridge }effectively solve point-to-point protocol integration, but they are inherently brittle where each new protocol pair necessitates a bespoke bridge implementation. KG-MAS circumvents this by creating a universal abstraction layer. Instead of relying on hard-coded translation logic, it uses the KG to semantically describe each component’s data formats and interaction patterns. This not only improves extensibility but also offers practical implementation for abstract models such as RAMI 4.0, which defines conceptual layers but lacks an executable framework.

Finally, KG-MAS’s semantic, agent-based architecture supports \textbf{Scalability \& Flexibility} inherently. Unlike traditional Knowledge-Based Systems, which enable reasoning but lack integrated execution or agent orchestration, KG-MAS embeds decision-making capabilities directly within an operational multi-agent system. By leveraging the KG as a blueprint for automatic agent generation, the integration of new physical or digital assets becomes seamless by requiring only a semantic description, not a full software redevelopment. This cohesive blend of semantic modeling, autonomous agents, and modular system design results in a CPS integration framework that is \textbf{more intelligent, adaptive, and maintainable}, fully aligned with the dynamic requirements of Industry 4.0.
\newpage
\section{Conclusion}
This project addresses the significant challenge of coupling heterogeneous physical and digital robotic environments within modern Cyber-Physical Systems. The proposed KG-MAS solution, is centered on the synergy between a centralized Knowledge Graph and a distributed Multi-Agent System. By synthesizing the strengths of disparate methodologies and mitigating their weaknesses, KG-MAS provides a holistic and powerful integration framework.

The core contribution of this approach lies in its semantic, model-driven architecture. By employing a KG as a dynamic and shared world model, the system moves beyond mere data exchange to enable true semantic interoperability. This allows autonomous agents to intelligently coordinate their actions based on a unified, context-rich understanding of the entire environment. A key advantage of this solution is the automatic agent generation process which builds agent logic directly from semantic descriptions. This significantly reduces development overhead and enhances system scalability, offering a distinguished alternative to brittle, custom-coded middleware bridges and the rigid master-control logic of traditional co-simulation.

In short, the synergy of a semantic world model, autonomous agents, and a modular design foundation creates a system that is not only more intelligent and coordinated but also significantly easier to maintain, extend, and adapt to the evolving demands of complex Industry 4.0 environments.

\clearpage
\pagestyle{empty}

\end{document}